# Validity of the one-dimensional limp model for porous media

*Olivier DOUTRES, Nicolas DAUCHEZ, Jean-Michel GENEVAUX, Olivier DAZEL*

LAUM, CNRS, Université du Maine, Av. O. Messiaen, 72095 LE MANS, FRANCE
nicolas.dauchez@univ-lemans.fr

**Abstract.** *A straightforward criterion to determine the limp model validity for porous materials is addressed here. The limp model is an "equivalent fluid" model which gives a better description of the porous behavior than the well known "rigid frame" model. It is derived from the poroelastic Biot model assuming that the frame has no bulk stiffness. A criterion is proposed to identify the porous materials for which the limp model can be used. It relies on a new parameter, the Frame Stiffness Influence FSI based on porous material properties. The critical values of FSI under which the limp model can be used, are determined using a 1D analytical modeling for a specific boundary set: radiation of a vibrating plate covered by a porous layer.*

## Keywords

Porous media, fibrous, limp model, Biot model, criterion.

## 1.Introduction

In recent years, poroelastic numerical models using finite element method have been widely developed to improve the acoustic efficiency of porous materials used in aeronautic and automotive industries. Classical methods use the Biot theory [1,2] to account for the displacements of both solid and fluid phases. To model three dimensional applications, six or four degrees-of-freedom per node are required depending on the chosen variable formulation [5,6].These numerical methods allow to predict the structural and fluid couplings induced by the poroelastic medium without any cinematic or geometrical assumptions. However, for large size finite element models, these methods can require a significant computational time.

To overcome this limitation, one can consider that the porous layer behaves like a dissipative fluid. Two porous "one-wave" formulations can be found: (i) the "rigid frame" model assumes that the solid phase remains motionless[2], (ii) the "limp" model assumes that the stiffness of the solid phase is zero but takes into account its inertial effects [8,9,10,11,12]. Because the motion of the solid phase is considered in the limp model, this model has to be preferred for most of the applications as in transports for example (car, train, aircraft), where the porous layers are bonded on vibrating plates. However, it is valid since the frame "flexibility" of the porous material has little influence on the vibroacoustic response of the system.

In a preceding paper [11], a criterion was proposed to identify the porous materials and the frequency bands for which the limp model can be used according to the boundary conditions applied to the layer. The identification process is based on a parameter, the Frame Stiffness Influence (FSI), determined from the properties of the porous material. This parameter, developed from the Biot theory [1,2] quantifies the intrinsic influence of the solid-borne wave [2] on the displacement of the interstitial fluid and is frequency dependent. In this study, the parameter FSI was compared to critical values obtained for different boundary conditions and porous thicknesses to give an estimation of the frequency bands for which the limp model can be used.

In this paper, the identification process is more straightforward to give a first estimation on the accuracy of using the limp model in the whole frequency range. It is based on a frequency independent parameter FSIr derived from FSI. Critical values of FSIr above which the limp model cannot be used are determined for porous materials of thicknesses from 1 to 5 cm and for a specific boundary condition set (see Fig.3). Here the sound radiation of a porous layer backed by a vibrating wall is presented.

## 2. Porous material modeling

### 2.1. Biot theory

According to Biot theory, three waves propagate in a porous media: two compressional waves and a shear wave. In this work, the applications are one dimensional and only the two compressional waves are considered. The motion of the poroelastic medium is described by the macroscopic displacement of solid and fluid phase, respectively denoted $u^s$ and $u^f$. Assuming a harmonic time dependence, the equation of motion can be written in the following form [11]:

$$-\omega^2 \frac{\tilde{\rho}_{12}}{\phi}\tilde{\Gamma}u^s - \omega^2 \frac{\tilde{\rho}_{22}}{\phi}\tilde{\gamma}u^f = \widehat{P}\nabla^2 u^s, \quad (1)$$

$$-\omega^2 \tilde{\rho}_{12} u^s - \omega^2 \tilde{\rho}_{22} u^f = \tilde{Q}\nabla^2 u^s + \tilde{R}\nabla^2 u^f, \quad (2)$$

with

$$\tilde{\Gamma} = \phi \left( \frac{\tilde{\rho}_{11}}{\tilde{\rho}_{12}} - \frac{\tilde{Q}}{\tilde{R}} \right), \quad \tilde{\gamma} = \phi \left( \frac{\tilde{\rho}_{12}}{\tilde{\rho}_{22}} - \frac{\tilde{Q}}{\tilde{R}} \right). \quad (3)$$

The tilde symbol indicates that the associated physical property is complex and frequency dependent. The inertial coefficients $\tilde{\rho}_{11}$ and $\tilde{\rho}_{22}$ are the modified Biot's density of the solid and fluid phase respectively. The inertial coefficient $\tilde{\rho}_{12}$ accounts for the interaction between inertial forces of the solid and fluid phases together with viscous dissipation. In Eq.(1,2), $\hat{P}$ is the bulk modulus of the frame in vacuum

$$\hat{P} = \frac{E(1+j\eta)(1-\nu)}{(1-2\nu)(1+\nu)}, \quad (4)$$

with $E$ the Young modulus, $\eta$ the loss factor, $\nu$ the Poisson ratio of the frame, $\tilde{R}$ is the bulk modulus of the fluid phase, $\tilde{Q}$ quantifies the potential coupling between the two phases and $\phi$ is the porosity.

In the considered geometry, the displacement of each phase is due to the propagation of two compressional waves traveling in both directions. They can be written in the form

$$u^s(x) = X_1 + X_2, \quad (5)$$
$$u^f(x) = \mu_1 X_1 + \mu_2 X_2, \quad (6)$$

where $X_i = S_i \cos(\delta_i x) + D_i \sin(\delta_i x)$ is the contribution of each compressional wave $i=1,2$, $S_i$ and $D_i$ being set by the boundary conditions. These waves are characterized by a complex wave number $\delta_i$ ($i=1,2$) and a displacement ratio $\mu_i$. This ratio indicates in which medium the waves mainly propagate. Here, the wave with the subscript $i=1$ propagates mainly in the fluid phase and is referred to as the "airborne" wave. The wave with the subscript $i=2$ propagates mainly in the solid phase and is referred to as the "frame-borne" wave.

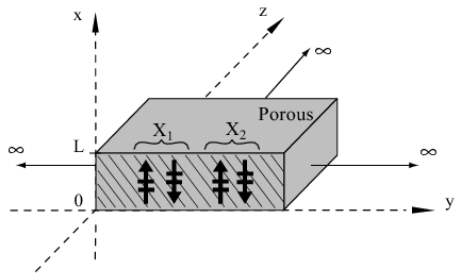

**Fig. 1.** One-dimensional porous modeling

## 2.2. Limp assumption

The limp model is derived from the Biot theory. It is based on the assumption that the frame has no bulk stiffness [8,9,10,11,12]: $\hat{P} = 0$. It is likely associated to "soft" materials like cotton and glass wool. This model describes the propagation of one compressional wave in a medium that has the bulk modulus of the air in the pores and the density of the air modified by the inertia effect of the solid phase and its interaction with the fluid phase.

Hence, by considering the assumption $\hat{P} = 0$ in Eq.(1), one gets a simple relation between the displacements of both solid and fluid phases. Then, substituting the solid displacement in Eq.(2) gives the propagation equation on $u^f$

$$\tilde{K}_f \nabla^2 u^f + \omega^2 \tilde{\rho}_{limp} u^f = 0, \quad (7)$$

with $\tilde{K}_f$ the bulk modulus of the air in the pores and $\tilde{\rho}_{\lim p}$ the modified density of the air. Expression of these coefficients can be found in reference [11,12].

## 3. Frame stiffness influence

The aim of this section is to propose a parameter based on the properties of the porous material which quantifies the influence of the frame stiffness on the porous behavior. This parameter is called FSI for Frame Stiffness Influence.

### 3.1. Development of the frequency dependent parameter FSI

The use of the limp model is possible when the contribution of the frame-borne wave is negligible in the considered application. This approximation implies in the expressions of the solid and fluid displacements (Eq.(5,6)) that:

- the contribution of the airborne wave $X_1$ is great compared to the contribution of the frame-borne wave $X_2$; this condition depends mainly on the boundary conditions : one configuration will be presented in section 4 to set critical values of the FSI parameter,

- considering the fluid motion (Eq.(6)), the displacement ratio $\mu_1$ associated to the airborne wave is great compared to the displacement ratio $\mu_2$ associated to the frame-borne wave: $\mu_2 / \mu_1 \ll 1$; this condition is independent from the boundary conditions and will be used to build the FSI parameter.

Hence, the FSI parameter is based on the assumption that the use of the limp model is possible when the frame-borne wave contribution is negligible in the considered application. The associated condition, $\mu_2/\mu_1 \ll 1$, can be written in terms of a frequency dependent parameter, FSI, expressed as a ratio of two characteristic wave numbers [11]

$$FSI = \frac{\delta_{limp}^2}{\delta_c^2} = \frac{\tilde{\rho}_{limp}}{\tilde{\rho}_c}\frac{\hat{P}}{\tilde{K}_f} . \quad (8)$$

$\delta_{\lim p} = \omega\sqrt{\tilde{\rho}_{\lim p}/\tilde{K}_f}$ is the wave number derived from the limp model and $\delta_c = \omega\sqrt{\tilde{\rho}_c/\hat{P}}$ is the wave number of a wave, called "c" wave, that propagates in a medium that has the bulk modulus of the frame in vacuum and the density of the frame in fluid

$$\tilde{\rho}_c = \rho_1 - \tilde{\rho}_{12}/\phi , \quad (9)$$

with $\rho_1$ the mass density of the porous material.

Figure 2 presents the FSI for the two characteristic material B and C [11]. Material B is a high density fibrous material and material C is a polymer foam with a stiff skeleton and a high airflow resistivity. The properties of these materials presented in Table 1 have been measured in our laboratory.

| Porous | B | C |
|---|---|---|
| Air flow resistivity: $\sigma$ (kN s/m$^4$) | 23 | 57 |
| Porosity: $\phi$ | 0.95 | 0.97 |
| Tortuosity: $\alpha_\infty$ | 1 | 1.54 |
| Viscous length: $\Lambda$ ($\mu$m) | 54.1 | 24.6 |
| Thermal length: $\Lambda'$ ($\mu$m) | 162.3 | 73.8 |
| Frame density: $\rho_1$ (kg/m$^3$) | 58 | 46 |
| Young's Modulus at 5 Hz: $E$ (kPa) | 17 | 214 |
| Structural loss factor at 5 Hz: $\eta$ | 0.1 | 0.115 |
| Poisson's ratio: $\nu$ | 0 | 0.3 |

**Tab. 1.** Measured properties of materials B and C.

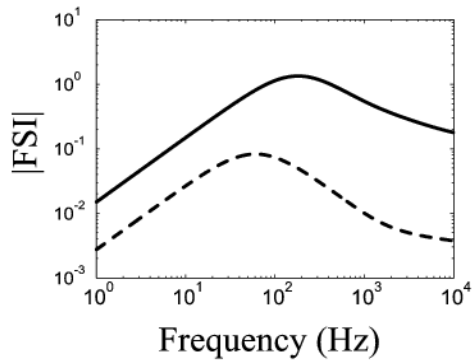

**Fig. 2.** FSI of material ( - -) B and (-) C.

This figure shows that the FSI parameter has a bell shape which amplitude increases with the bulk modulus of the porous skeleton. The maximum amplitude occurs at the decoupling frequency defined by Zwikker and Kosten [13]:

$$f_{ZK} = \frac{\phi^2 \sigma}{2\pi\rho_1} . \quad (10)$$

This frequency indicates the frequency below which the viscous forces on the material are superior to the inertial forces per unit volume. It is generally used to determine the critical frequency above which an acoustical wave propagating in the fluid phase would not exert a sufficient force to generate vibrations in the solid phase.

### 3.2. A simplified frequency independent parameter FSIr

The main objective of the paper is to propose a straightforward identification process which is more easy to carried out compared to the one presented in ref[11]. The criterion proposed in this paper consists in comparing a frequency independent parameter which characterizes the frame influence with critical value. This frequency independent parameter is set as the maximum value of FSI to ensure the uniqueness of the solution in the whole frequency range. Thus, as mentioned previously, it can be approached from the mass densities of both the limp and the "c" waves expressed at the frequency $f_{ZK}$.

Assuming that the density of air $\rho_f$ is negligible compared with the one of the porous material $\rho_1$, these densities are given by

$$\tilde{\rho}_c(f_{ZK}) \approx \rho_1(1 - j\frac{1}{\phi}) , \quad (11)$$

$$\tilde{\rho}_{limp}(f_{ZK}) \approx \rho_1\frac{(1 - j\phi)}{(1 + \phi^2)} . \quad (12)$$

Hence, the modulus of the maximum FSI at $f_{ZK}$ is given by

$$\text{FSI}_r = |\text{FSI}(f_{ZK})| \approx \frac{\hat{P}}{P_0}\frac{\phi}{1 + \phi^2} . \quad (13)$$

FSIr is then easy to calculate and requires the measurement of the bulk modulus of the skeleton $\hat{P}$ and the porosity ($\phi$). The two parameters FSIr and $f_{ZK}$ are given in Table 2 for materials B and C.

| Material | B | C |
|---|---|---|
| $f_{ZK}$ (Hz) | 57 | 186 |
| FSI$_r$ at $f_{ZK}$ | 8.42 10$^{-2}$ | 1.43 |

**Tab. 2.** Simplified FSI parameter of materials B and C.

## 4. Determination of critical FSI values

In the previous section, the simple parameter FSIr based on the physical properties of the material has been introduced. The next step is to identify, for a specific boundary condition set, the critical values of FSI under which the limp model can be used instead of the Biot model. These critical values are determined from the difference between the limp and the Biot model carried out for a wide range of acoustic materials: hence, the critical FSI value is independent of the tested material.

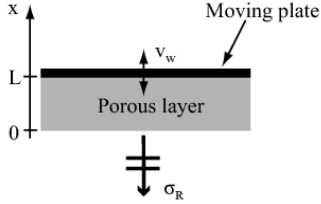

**Fig. 3.**     Sound radiation of a porous layer backed by a vibrating wall.

The chosen configuration is presented in Fig.3. The porous layer is excited by a vibrating plate at $x = L$ and radiates in an infinite half-space at $x = 0$. This configuration corresponds to trim panels, cars roofs or airplane floors. The radiation efficiency factor $\sigma_R$, defined as the ratio of the acoustic power radiated $\Pi_a$ over the vibratory power of the piston $\Pi_v$, is used as vibroacoustic response:

$$\sigma_R = \frac{\Pi_a}{\Pi_v} = \frac{p(0)v^*(0)}{\rho_f c_f v_w^2} \ . \qquad (14)$$

A vibrating surface area of 1 m² is considered here. Boundary conditions associated to this configuration are [14]: continuity of stress and total flow at $x = 0$. At $x = L$, the velocity of the fluid and the velocity of the frame are both equal to the wall velocity

$$j\omega u^s(L) = j\omega u^f(L) = v_w \ . \qquad (15)$$

The vibroacoustic response is derived using the Transfer Matrix Method (TMM)[2]. This method assumes the multilayer has infinite lateral dimensions and uses a representation of plane wave propagation in different media in terms of transfer matrices. To ensure a one-dimensional representation, the multilayer is excited by plane waves with normal incidence. The porous layer is either simulated using the Biot model or the limp model presented in section 2. Fig.4 show the Biot and limp simulations of the radiation efficiency of materials B and C of thickness 2 cm. For both materials, an increase of the radiation efficiency is observed around the first $\lambda/4$ resonance frequency of the frame: around 200 Hz for material B and 1000 Hz for material C.

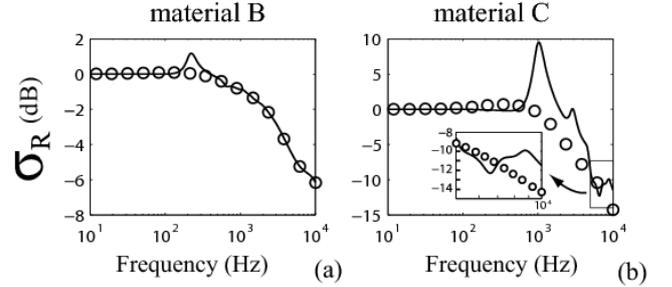

**Fig. 4.**     Radiation efficiency simulated with the Biot model (solid line) and the limp model (circles): (a) material B, (b) material C.

To determine the critical FSI value, the difference between the two models is derived by the absolute value of the difference of the two responses $\Delta\sigma_R = \left|\sigma_{R(Biot)}\right| - \left|\sigma_{R(\lim p)}\right|$. The maximum accepted difference between the two models is set to 3 dB and corresponds to a classical industrial demand. In order to determine a critical FSI value independent of the tested material, the difference between the two simulations is plotted as a function of the frequency dependent parameter FSI for a wide variety of porous materials (256 simulated materials). The critical FSI value corresponds to the minimum FSI value for which the model difference exceeds the maximum acceptable value of 3 dB [11].

The abacus given in Fig.5 present the minimum FSI critical values determined for 5 different porous thicknesses. For a given material, the limp model can be used if its FSIr is situated below the critical value (white area of the abacus) and the Biot model should be preferred if FSIr exceeds the critical value (gray area of the abacus).

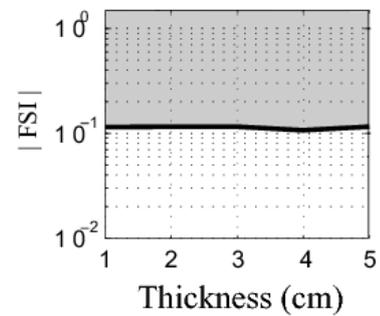

**Fig. 5.**     Evolution of FSI critical value as function of the porous thickness.

## 5. Discussion and conclusion

A straightforward method is proposed to determine if the limp model can be used in the whole frequency range (1-10000 Hz). The procedure is as follows:

- Two properties of the porous materials, $\hat{P}$ and $\phi$, have to be measured. (see TAB.1 for materials B and C),
- The parameter FSIr is evaluated using Eq.(13).
- The critical values of FSI are chosen in Fig.5 according to the thickness of the porous layer.
- FSIr is finally compared to the critical values: the limp model can be used in the whole frequency range if FSIr is below the FSI critical value.

In the case of material C, FSIr is equal to 1.4 (see TAB.2) which is above the FSI critical values of the radiation configuration and for all thicknesses: the Biot model should be preferred for all layer thicknesses. The FSIr of material B is equal to $8.4 \cdot 10^{-2}$ (see TAB.2) which is below the FSI critical values of the radiation configuration for all thicknesses: the limp model can be used for all porous thicknesses. These predictions agree with the simulations presented in Figs.4. Note that for material B, the increase of the radiation efficiency induced by the frame motion do not exceed the maximum accepted difference between the Biot and limp modelizations of 3dB.

The proposed method is easy to carry out and allows to estimate if the one-dimensional limp model can be used instead of the complete Biot model without making any numerical simulations of the configuration nor experimental studies. Note that the use of the limp model can be particularly interesting in order to decrease the computational time for large finite element calculations which include porous materials. The criterion method has been presented here in the case of the radiation efficiency of a plate covered by a porous layer of different thicknesses. It has been shown that the prediction of the material for which the limp model can be used is in close agreement with 1D simulations.

## Acknowledgements

This study was supported in the scope of the CREDO research project co-funded by the European Commission.

## References

[1] BIOT, M.A.: *The theory of propagation of elastic waves in a fluid-saturated porous solid I. Low frequency range; II Higher frequency range*. J. Acoust. Soc. Am. **28**, p. 168-191, 1956.

[2] ALLARD J. F.: *Propagation of Sound in Porous Media: Modelling Sound Absorbing Materials*, Elsevier Applied Science, London, 1993.

[3] PANNETON R., ATALLA N.: *An efficient scheme for solving the three-dimensional elasticity problem in acoustics*, J. Acoust. Soc. Am. **101**(6), p. 3287–3298, 1998.

[4] HÖRLIN N. E., NORDSTRÖM M.., GÖRANSSON P.: *A 3-D hierarchical FE formulation of Biot's equations for elastoacoustic modeling of porous media*, J. Sound. Vib. **254**(4), p. 633–652, 2001.

[5] ATALLA N., PANNETON R., DEBERGUE P.: *A mixed displacement-pressure formulation for poroelastic materials*, J. Acoust. Soc. Am. **104**(4), 1444-1452, 1998.

[6] DAUCHEZ N., SAHRAOUI S., ATALLA N.: *Convergence of poroelastic finite elements based on Biot displacement formulation*, J. Acoust. Soc. Am. **109**(1), 33–40, 2001.

[7] RIGOBERT S., ATALLA N., SGARD F. : *Investigation of the convergence of the mixed displacement pressure formulation for three-dimensional poroelastic materials using hierarchical elements*, J. Acoust. Soc. Am. **114**(5), 2607–2617, 2003.

[8] BERANEK L. L.: *Acoustical properties of homogeneous, isotropic rigid tiles and flexible blankets*, J. Acoust. Soc. Am. **19**(4), p. 556-568, 1947.

[9] INGARD K. U.: *Notes on sound absorption technology*, Noise Control Foundation, New York, 1994.

[10] DAZEL O., BROUARD B., DEPOLLIER C., GRIFFITHS S.: *An alternative Biot's displacement formulation for porous materials*, J. Acoust. Soc. Am. **121**(6), 3509-3516, 2007.

[11] DOUTRES O., DAUCHEZ N., GENEVAUX J.M., DAZEL O.: *Validity of the limp model for porous materials: A criterion based on Biot theory*, J. Acoust. Soc. Am. **122**(4), 2038-2048, 2007.

[12] PANNETON R.: *Comments on the limp frame equivalent fluid model for porous media*, J. Acoust. Soc. Am. **122**(6), EL 217-222, 2007.

[13] ZWIKKER C., KOSTEN C.W.: *Sound absorption materials*, Elsevier applied science, New York, 1949.

[14] DOUTRES O., DAUCHEZ N., GENEVAUX J.M.: *Porous layer impedance applied to a moving wall: Application to the radiation of a covered piston*, J. Acoust. Soc. Am. **121**, 206-213, 2007.